# Analysis of SVN Repositories for Remote Access

Sadaf Solangi and Safeeullah Soomro
Department of Computing and Technology
Faculty of Engineering & Technology, Indus University
Karachi, Pakisstan
e-mail:{sadaf.solangi,ssoomro}@indus.edu.pk

Suhni Abbasi
Institute of Information Technology Center
Faculty of Social Sciences, Sindh Agricultural
University Tando Jam, Pakistan
Sabbasi11@gmail.com

*Abstract—* Software Evolution is considered to be essential and challenging characteristic in the field of software engineering. Version control system is an incremental versions tracking system, introduced to avoid unnecessary overwriting of files such as programming code, web pages and records. It also helps to decrease the confusion affected by duplicate or outdated data. In this proposed research SVN repository is maintained and analyzed for msitone.wikispaces.com to minimize the efforts as well as resources for the future users. We have used two semester data for the analysis purpose that is observed SVN repository. The result shows that, implementing the SVN repositories are helpful for maintenance of the Wikispaces as it also reduce the cost, time and efforts for their evolution. Whereas without implementing the SVN repositories Wikispaces were just supposed to be building the house by putting each brick from start.

## I. INTRODUCTION

Source control (version control) is the central component of modern software development process. Various kinds of version control systems are available like, CVS (Concurrent Version System), SVN (Subversion), Git, Mercurial, Bazaar, LibreSource, and Monotone. Open source Concurrent Versions System (CVS) and subversion (SVN) VCSs were mainly designed to work with plain text files that were also control other file data type[6].SVN is a version control system, introduced to avoid unnecessary overwriting of files and decreases the confusion affected by duplicate or outdated data. CollabNet Inc developed revision control system in 2000 [1]. used to maintain historical versions files as well as current files like, source code, Webpages and documentation. This open source revision control system used to perform changes in text based files i.e. source code. SVN performs work on directories, files helpful to manage as well change them. This feature of SVN provide the facility to return over the prior version of your program at any point which were built at any time. If any user enter the wrong code unintentionally and missed all the chances to undo the code, the versioning control system can resume the code from any point.
Described that during the 2005 semester course they as a team showed their four open-source personal research collections [10].How digital repositories were well used and for what, by address of recording [7].

### A. SVN Repository

A repository is information about the database that is shared engineered artifacts created and used by enterprise. A common repository allows instruments to share information, not including a common Repository, and it will need a particular protocol Exchange of information between machines [13].Examples of such sample includes, software, documents etc. Repository has three types of sessions such as trunk, branches and tags. Trunks contain latest development code, Branches possess multiple versions of the same product lines for the development can be used to produce and Tags are the any version are released to the public, with source code version number tag. The MSR(Mining Software Repository) launches a significant challenge for both investigators and experts in the field of software repository mining [11].

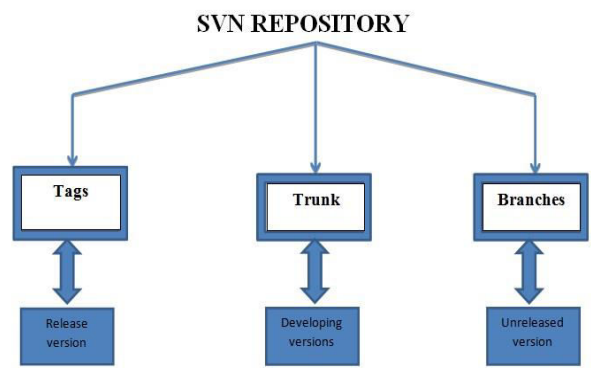

Figure1. General Concept of SVN repository

### B. Visual SVN Server and Tortoise SVN Client

Visual SVN Server provides facility to manage, install multi-functional subversion server easily on windows platform. Tortoise SVN is used as client tool for maintaining the SVN repository by applying their multiple functions and commands. It manages the projects that are located in the subversion repository. Visual SVN Server is used as central and local repository.

The user in order to get himself facilitated has to perform certain tasks. The author creates the user accounts on the Visual SVN Server for remote access of data. Then the link is being created. After that, the repository is created which is being linked through the checkout operation i.e. the Visual SVN Server (Server Side) is now linked to the Turtoise SVN tool (Client side).User now has the authentication to access the data and can analyze the data by using different modification operations. Every time, the data is modified a new version is being created online. The modification results of different authors/users can be analyzed in the graphical or statistical form. The user can latter on access the data remotely by using Visual SVN Server link and user account.

Software repositories for analyzing received a lot of attention from researchers in recent years have been to, this first version control and issue tracking systems to extract the raw data is mandatory [5].

## 2 SVN Repository Analysis

### C. SVN Design for wiki

Subversion is an open source (versions control system), which is used for managing the up-to-date old versions data like coding, website pages, and records. In the proposed research, SVN repository was maintained and analyzed for msitone.wikispaces.com, to minimize the efforts as well as resources for the future users. Software world takes the large software repositories for example source Forge (350,000+ projects), GitHub (250,000+ projects),and Google Code(250,000+projects)[2]. Msitone.wikispaces.com page was designed for MSIT 2k12 batch of ITC Tandojam and two semesters' data i.e. 01-Jan 2013 to 31-Dec 2013 from designed wikispace page was analyzed in this study. This study focused on the reuses of web pages and the data for the next batch of MSIT as an evolution phase. For this purpose archive of MSIT page was required and visual SVN server also used provide the remote and local access to the manually stored archives of wikispaces to the version control users. Tortoise SVN tool was used as SVN client. Visual SVN server provides the interface between SVN client and archives of wikispace. Then analyzed wiki data from visual SVN Server by using different actions of Tortoise SVN tool and revisions automatically update on server.

Subversion is essentially a third-generation version management system is a super up example [4].ROSE tool that is applied to version records to data mining that monitors programs besides associated variations [12].

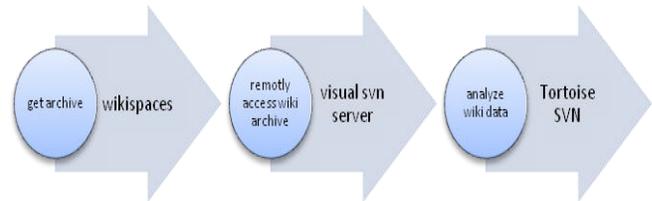

Figure 2.1 Conceptual diagram for SVN to wiki relation

### D. Visual SVN Server

Visual SVN Server provides facility to manage, install multi-functional subversion server easily on windows platform. It is allocated as a separate installation set with the latest versions of all the required components. Visual SVN Server uses the built-in isolated violations over HTTP to communicate with clients and web browsers Apache HTTP Server.

### E. Tortoise SVN Client

Tortoise SVN is free software used by developers to manage different versions of the source code. Tortoise SVN tool is used to maintain the SVN repository by applying their multiple functions and commands. It manages the projects that are located in the, a subversion repository. Tortoise SVN repository have three types of sessions such as tag, branches, and trunk, which are discussed here in detail.

### F. Tag

Tags are used to highlight the versions in the repository history and tag must be created before releasing any version.

### G. Branches

Branches are used to create the development lines for multi-versions of the same product so as that a storage place to fix bugs in the stable release. It also provided side-line development.

### H. Trunk

Trunks are the core leading line of development in SVN repository.

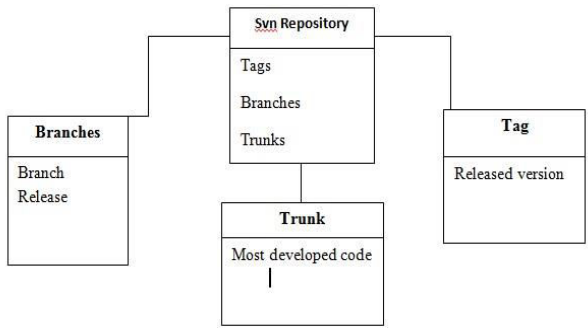

Figure 2.2 SVN repository schemas

*I.   SVN-WIKI architecture*

The research process was conducted through SVN-Wiki Architecture. This architecture aims provides the complete working flow in sequence of step. The architecture is shown in Figure 2.3

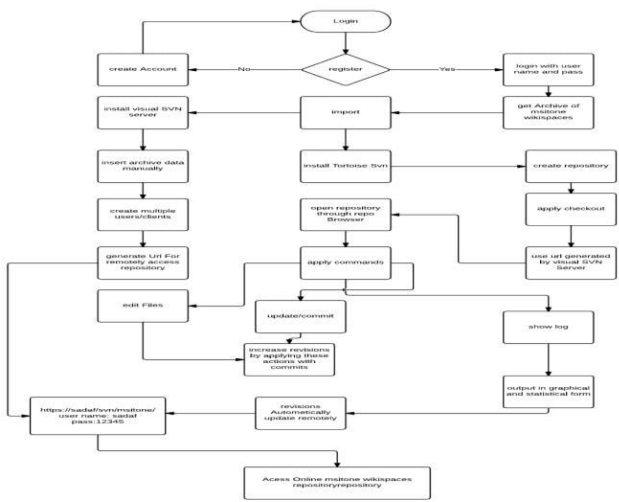

Figure 2.3 SVN-WIKI whole Process Architectural Diagram

## 3  Results

*J.   Populate the archives of the wikispace data over server using SVN server*

Msitone wikispaces information was collected in the form of Archive. It was then manually insert whole data into Visual SVN Server so as to remotely access on data. Wiki data managed in Visual SVN Server tool is shown in Figure 3.1.

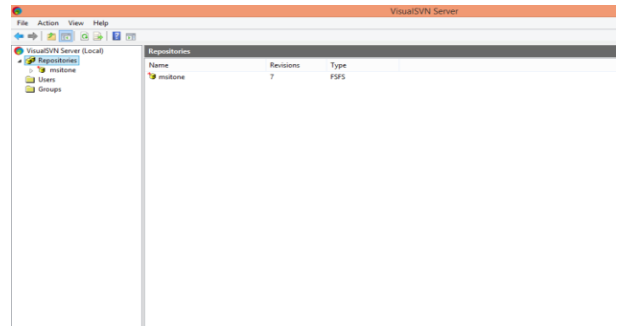

Figure 3.1 Insert manually data into Visual SVN Server

*K.   Create user/client for remote access*

Visual SVN Server was not allowed to Access your data without user account. Create user and then easily access the wiki spaces information remotely. Through this https://sadaf/svn/msitone/ and username: sadaf password: xxxxx.

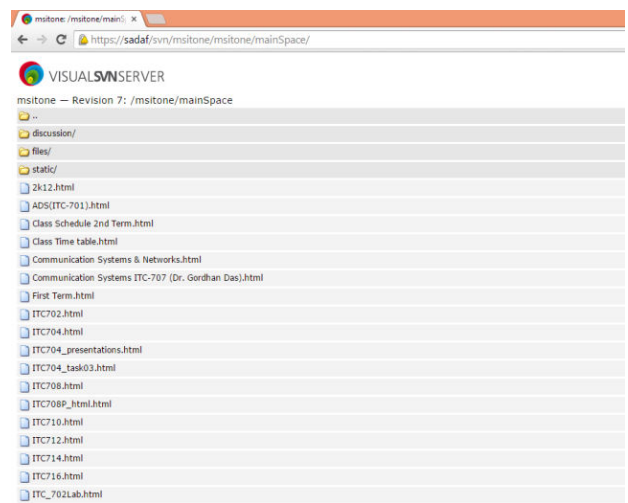

Figure 3.2 Remotely Access data from server

*L. Accessing repository through Tortoise SVN Tool*

Tortoise SVN Client provides facility to create repository for historical data. Process was starting to create SVN repository to analyze wikispace information and increase revisions by different authors.

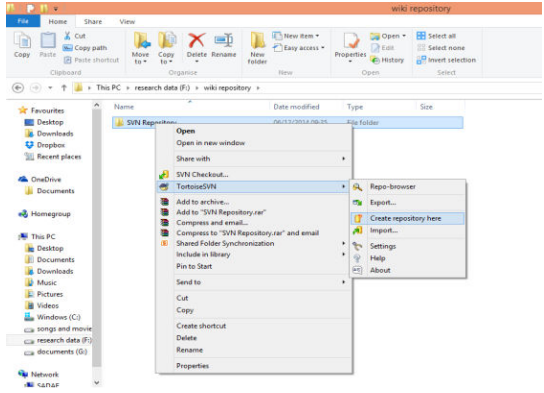

Figure 3.3 Process of creating SVN Repository By using Tortoise SVN Client Tool

*M. Checkout repository*

Checkout is a very important phase in this whole process. The checkout is applied on SVN repository using URL of data which were accessible through Visual SVN Server and established link/interface between both tools so that SVN repository can easily be analyzed.

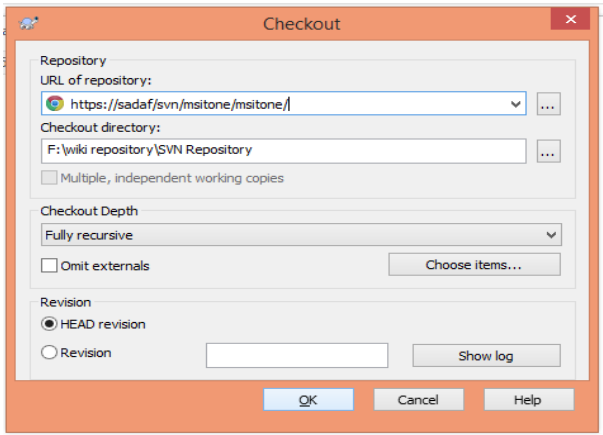

Figure 3.4 Apply checkouts On SVN Repository and give the server URL

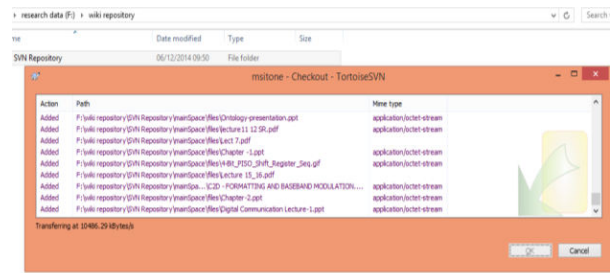

Figure 3.5 specify that checkout process after providing the username and password

*N. Analyze files, directories through different actions.*

Analyzing phase was started; Go repo browser and then applied more actions of Tortoise SVN tool on that data and important thing was that the changes were automatically updated on server.

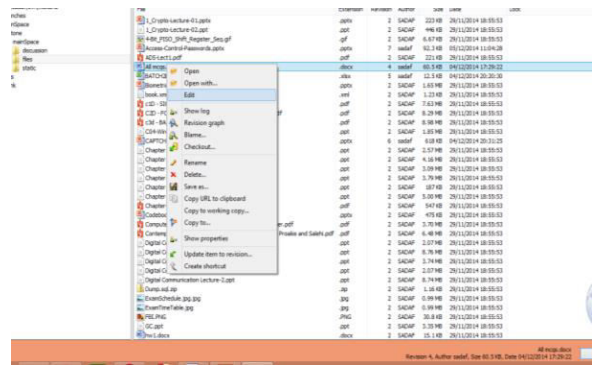

Figure 3.6 specifies the checkout process after authentication

*O. Graphical and statistical results of revisions and commits by authors.*

Tortoise SVN client provides graphical and statistical results of all activities which were applied on that data. Results based on commits, authors, date and files etc.

*P. Statistical results:*

Statistical results contain total counts of revisions, commits, files, authors and most, least users. These results are generated with Tortoise SVN. One of the sample results is described in Table 3.1 and 3.2.

| Ranking Number | Authors | Files |
|---|---|---|
| 1 | Sadaf Solangi | All mcqs.docx |
| 2 | Erum | BATCH2K12-LIST.xlcx |
| 3 | Gulshan | 1-crypto-lecture-02.pptx |
| 4 | Haseena | 4-bit-PISO-Shift-Register-Seq.gif |
| 5 | Mehtab | ExamSchedule.jpg.jpg |
| 6 | Nareena | ADS-Lect1.pdf |
| 7 | Bushra Tahira | Digital Communication Chapter-2.ppt |
| 8 | Noreen | Biometrics.pptx |
| 9 | Paras | CAPTCHA-and-Firewalls.pptx |
| 10 | Suhni Abbasi | C2D - FORMATTING AND BASEBAND MODULATION.pdf |

**Table 3.1 Top 10 files and 10 hot Author List**

| Analyzed Actions | Results |
|---|---|
| First revision number | 1 |
| Total revision number | 50 |
| Total file revision count | 247 |
| Author count | 15 |
| First revision date | Sep 22, 2014 06:37pm |
| Last revision date | Nov 26, 2014 09:39am |
| Most active author | Sadaf Solangi |
| Least active author | Sehrish |
| Particular weeks count | 4 |

**Table 3.2 Result of General statistical SVN Repository**

*Q. Graphically results:*

Graphical results consist of (revisions, commits, authors, date).

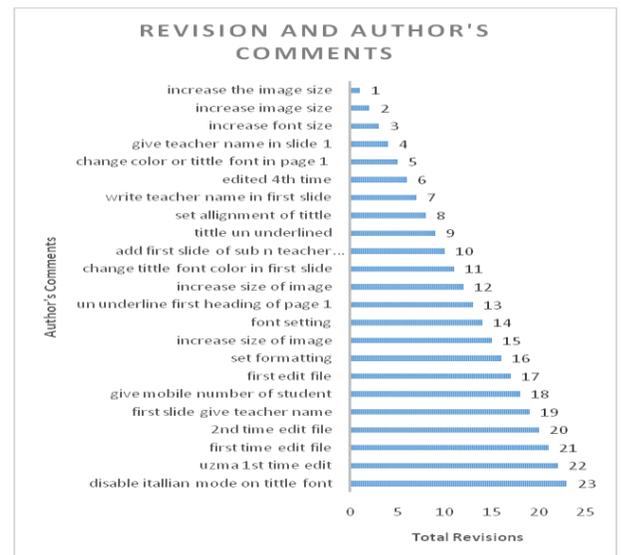

Figure 3.7 commits on revision and total revisions

Figure 3.7 shows all commits given by authors on different SVN repository files their particular dates as well as total number of revisions was also calculated.

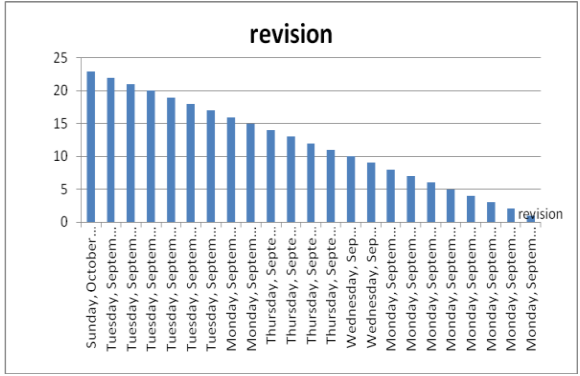

Figure 3.8 Revisions by dates

Figure 3.8 describes total revisions and every revision has been done on multiple dates.

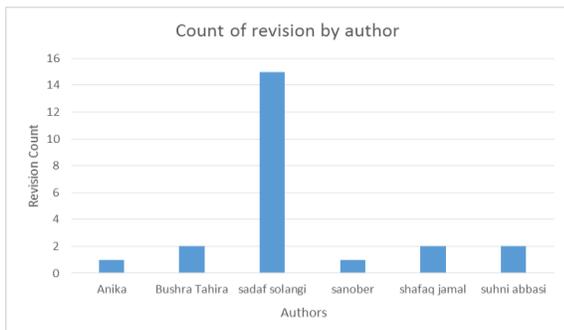

Figure 3.9 Authors and revision count

Figure 3.9 shows the results of different authors and number of percent by every authors has participated in the form e.g. edit any file, give the commits and also applied edit, delete and modify commands.

## 4  Conclusion

The study was focused on maintaining, analyzing and reusing the http://msitone.wikispaces.com repository. The data of academic year 2012-2013 was stored in the form of archives and backups. Reusing that data for future purpose needs extra efforts and resources. To overcome this problem the SVN repository was developed on archives of wiki created for 2K12 masters batch of Information Technology Centre.

In this research archive was downloaded from MSIT wikispaces page. The downloaded archive repository was created on client side and then it was stored on server. Multiple users and authors had accessed and analyzed that repository.

In this study SVN repository was analyzed by using Visual SVN Server and Tortoise SVN Client tools. SVN Tortoise Client tool was used for maintaining SVN repository by commands like commit, update, checkout, delete, revisions. Visual SVN Server was used for remotely storage purpose.

Manually editing of data was time consuming, if there is no backup of data, one cannot revert the changes at specific point.

The results were supported by German in which CVS was used to maintain files for software development [3]. The results had also showed similarity with [2] who used Boa, a domain specific language and infrastructure to testing MSR (Mining Software Repositories) which substantially decreases programming efforts.

The results were also supported by [10] that had managed institutional repository and analysed that repository was more useful than regular basis update of any academic record. Open access digital institutional repository of universities to be engaged in an exchange of scholars by using a relatively new technology[8].

## 5  Limitations

1. The users need to install both the Visual SVN Server and Tortoise SVN Client on their systems and create a repository itself.

2. Archived data need to be saved to avoid any mishandling.

**REFERENCES.**

[1] CollabNet. 2014. CollabNet Subversion.http://www.collab.net/news/press/ameritas-selects-collabnet-teamforge-its-agile-alm-platform-drive-cost-savings-and Last visited on: 3/16/2014.

[2] R. Dyer, H. A. Nguyen, H. Rajan, and T.N. Nguyen, "Boa: a language and infrastructure for analyzing ultra-large-scale software repositories", ICSE, 2013, pp.422-431.

[3] D.M. German, "Mining CVS repositories, the softChange experience: In Proc.Int'l Workshop on Mining Software Repositories", 2004, pp.17-21.

[4] J.V.Gurp, and C.Prehofer,Version management tools as a basis for integrating Product Derivation and Software Product Families. In Proc of the Workshop on Variability Management-Working with Variability Mechanisms at SPLC, 2006, pp. 48-57.

[5] S. Kim, T. Zimmermann, M. Kim, A. Hassan, A. Mockus, T. Girba,M. Pinzger,E. J. Whitehead, and A. Zeller, "TA-RE. An exchange language for mining software repositories". J. MSR, 2006, pp.22-25.

[6] C. Kussmaul, and M. College, "Supporting teams with open source software tools",J. NCIIA, 2008, Pp.141-147.

[7] D. Nicholas, I. Rowlands, A. Watkinson, D. Brown, and H. R Jamali, "Digital repositories ten years on: what do scientific researchers think of them and how do they use them?", J. Learned Publishing, 25(3), 2012,pp.195–206.

[8] P. Shields, N. Rangarajam, and l. stewart, "Open access digital repository sharing student research with the world", JPAE.18 (1), 2012, pp.157-181.

[9] S.K. Sowe, I. Samoladas, I. Stamelos, and L. Angelis, "Are FLOSS developers committing to CVS/SVN as much as they are talking in mailing lists? Challenges for integrating data from Multiple Repositories", J. WoPDaSD, 2008, Pp.49-54.

[10] M. Taufer, P.J.Teller, A. Kerstens, and R. Romero, "Collaborative Research Tools for Students, Staff, and Faculty",In Proc. of the Int'l SUN Conference on Teaching and Learning, 2007, Pp.1-6.

[11] L. Voinea, and A.Telea, "Mining Software Repositories with CVSgrab:", In Proc. of the 2006 Int'l Workshop on Mining Software Repositories, 2006, pp.167-168.

[12] T. Zimmermann, P. Weißgerber, S. Diehl, and A. Zeller, "Mining Version Histories to Guide Software Changes", IEEE Transactions on Software Engineering,31 (6), 2005,pp.429-445.

[13] P. A.Bernstein, and U.Dayal,"An Overview of Repository Technology", Conf VLDB,1994,Pp.705-713.

[14] A.D.Kingsley,"The advocacy and awareness imperative: a repository overview",conf(VALA), 2010,pp.1-13.